\documentclass{appolb}
\usepackage{graphicx}

\begin{document}
\title{VOXES, a new high resolution X-ray spectrometer for low yield measurements with diffused sources%
\thanks{Presented at $2^{nd}$ Jagiellonian Symposium on Fundamental and Applied Subatomic Physics}%
}

\author{~
\address{~}
\\
{A.~Scordo$^{1}$, H.~Shi$^{1}$, C. Curceanu$^{1}$, M. Miliucci$^{1}$, F. Sirghi$^{1,2}$, J.~Zmeskal$^{3}$
}
\address{$^{1}$INFN Laboratori Nazionali di Frascati, Frascati (Roma), Italy\\
$^{2}$Horia Hulubei National Institute of Physics and Nuclear Engineering (IFIN-HH), Magurele, Romania\\
$^{3}$Stefan-Meyer-Institut f\"ur subatomare Physik, Vienna, Austria\\}
}

\maketitle

\begin{abstract}
The VOXES project's goal is to realize the first prototype of a high resolution and high precision X-ray spectrometer for diffused sources, using Highly Annealed Pyrolitic Graphite (HAPG) crystals combined with precision position detectors.
The aim is to deliver a cost effective and easy to handle system having an energy resolution at the level of few eV for X-ray energies from about 2 keV up to tens of keV.
There are many applications of the proposed spectrometer, going from fundamental physics (precision measurements of exotic atoms at DA$\Phi$NE collider and J-PARC, precision measurement of the $K^-$ mass solving the existing puzzle, quantum mechanics tests) to synchrotron radiation and applications (X-FEL), astronomy, medicine and industry.
Here, the basic concept of such a spectrometer and the first results from a measurement of the characteristic Cu $K_{\alpha 1}$ and $K_{\alpha 2}$ X-ray lines are presented. 
\end{abstract}

\section{Introduction}
The possibility to perform very high resolution measurements of the X-rays emitted in various
processes is becoming a priority in many fields of fundamental science, from particle and nuclear
physics to quantum mechanics, as well as in astronomy and in many applications using the synchrotron
light sources or X-FEL beams, in biology, medicine and industry.
The X-ray detection systems are presently passing through a remarkable fast development process,
with new type of detectors being constantly realized, having steadily improved performances in terms of
efficiency, energy resolution and costs.
Silicon, and more generally solid state, detectors such as Silicon Drift Detectors (SDDs) or
Charged Coupled Devices (CCDs), have been intensively used for position and energy measurements.
Recently, their use as large area spectroscopic detectors has been explored in the framework of the
SIDDHARTA experiment \cite{SID1,SID2,SID3}, with a lot of success. However, when used for precision
transition lines width measurements these devices are limited by their intrinsic resolution related to the
Fano Factor \cite{FANO} and to the electronic noise. Such detectors
have resolutions at 6 keV energy of about 120 eV (FWHM) at best. In the past, experiments performed at
the PSI laboratory measuring pionic atoms \cite{PIAT} pioneered the possibility to combine these type of
detectors (CCDs) with crystals, to enhance the energy resolution using Bragg reflection principle. 
The used crystals were of silicon type and the energy range achievable to the system was limited to few keV, 
due to the crystal structure, which is also responsible for the intrinsic low efficiency of such crystals.
Other type of high resolution detectors being presently under development are the Transition Edge
Sensors (TES). The achievable energy resolution is excellent (few eV at 6 keV), but the active area is
still very small \cite{TES}. Moreover, the cost of these detectors is prohibitively high and their use rather
laborious, due to the large complex cryogenic system needed to reach the operational temperature of $\simeq\,50\,mK$.
In spite of the fast evolving radiation detectors market, the aim of performing sub-eV precision measurements 
not only of photons produced from well collimated and point-like sources
but for photons coming from extended and diffused sources as well, like the ones emitted in the exotic atoms radiative transitions, 
is still to be reached. 

\section{Wavelength Dispersive X-ray spectroscopy with HAPG} 

Standard X-ray spectrometers, both in the von Hamos \cite{VON1,VON2,VON3} or Johann \cite{JFIG} configuration, have already proved their excellent energy resolution but, unfortunately,
the intrinsic low efficiency and rigidity of the standard crystals made them unsuitable as possible detector candidates for diffused X-rays, 
especially when low yield processes have to be investigated.
In these conditions, a new possibility comes from the
development, in the last decade, of the Higly Annealed Pyrolitic Graphite mosaic crystals (HAPG,\cite{HAP1,HAP2,HAP3}),
consisting in a large number of
nearly perfect small pyrolitic graphite crystallites. The unique structure of such crystals enables them to be highly
efficient in diffraction in an energy range between 2 keV and several tens of keV. 
Mosaicity, defined as the FWHM of the small crystallites orientation distribution,
makes it possible that even for a fixed incidence
angle on the crystal surface, a photon within a certain energetic distribution can be reflected, because each photon of
this energetic distribution can find a crystallite plane at the right Bragg angle (\cite{MOSA}).
HAPG is an artificial graphite ($2d_{002} = 6.708 ${\AA}) with intrinsic properties including high reflectivity, efficiency and spatial resolution, 
which are already proved to be an ideal solution for focusing soft X-rays on position sensing detectors. 
The mosaicity is also responsible for the dramatic increase of the integrated reflectivity in comparison to perfect crystals in
an energy range between 2 keV and several tens of keV \cite{HAP2}.


\section{von Hamos and Johann spectrometers}

\begin{figure}[htb]
\centerline{%
\includegraphics[width=11cm]{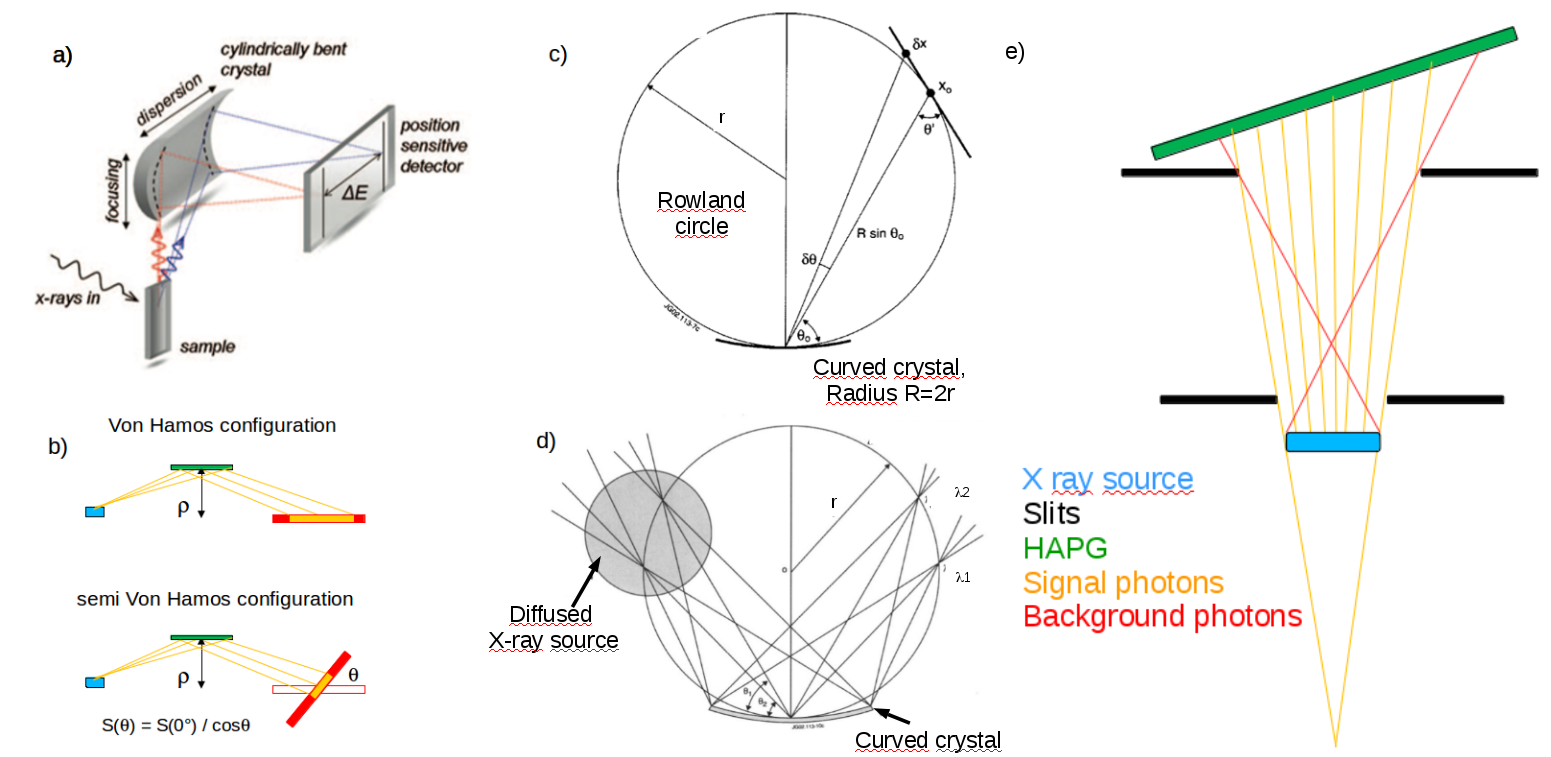}}
\caption{\em von Hamos (a \cite{VHFG},b) and Johann (c,d,\cite{JFIG}) spectrometer schemes. Simulated point-like source condition (e).}
\label{compare_big}
\end{figure}

The idea to combine the dispersion of a flat crystal with the focusing properties of cilindrically
bent crystals was put forward from von Hamos in 1933 \cite{VON1},
and has already proven its high-resolution capabilities \cite{VON2,VON3} and excellent focusing properties \cite{HAP2}.
However, in all these experiments the crystals have been used to focus very small and collimated
sources. In the past, an attempt to use silicon crystals in the Johann configuration for wider sources has been carried out,
where the pionic hydrogen and deuterium X-ray lines have been measured; however, these lines are in the narrow energy
range of 2.4-3.1 keV and those systems had a very low efficiency \cite{BEER}. 
With the advent of the HAPG, both the Johann and the von Hamos spectrometer could be again considered as possible candidates
for high resolution and efficiency detectors for diffused X-ray sources. 
A comparison of the two configurations is reported in fig. \ref{compare_big}.

\noindent The advantages of using the Johann configuration consist in having, together with an autoselective scheme in which a diffused source can be used 
(see fig. \ref{compare_big}d), typically smaller and more compact geometries with respect to the von Hamos one;
for a fixed $\rho$ the typical source-to-crystal and crystal-to-detector distances are much smaller than in the von Hamos case, resulting in a higher geometrical efficiency
(larger solid angle) and a lower X-ray absorption in air. The main disadvantage of such a scheme is related with the necessity, when a typical linear position detector is used, to
correct the measured spectra for non linear effects (see fig \ref{compare_big}c).
On the other hand, the advantages of using the von Hamos configuration (see fig. \ref{compare_big}a) consist in having a linear energy spectrum with a standard
position detector, in an energy range for fixed $\rho$ of the crystal given by the crystal and detector dimensions and in the vertical focusing. 
The greater source-to-crystal and crystal-to-detector distances result in a lower background contamination from other particles if used in high background environements like particle accelerator. 
In order to improve the dynamic range, a ``semi'' von Hamos configuration can also be used (see fig. \ref{compare_big}b) where the position detector is rotated of an angle $\theta$ with respect to the cilindrically 
bent crystal axis; in this particular configuration the illuminated surface portion of the position detector scales following the $S(\theta)=\frac{S(0)}{cos(\theta)}$ equation, 
being $\theta$ the angle by wich the detector is rotated (see fig. \ref{compare_big}b) and $S(\theta)$ the illuminated portion of the detector surface at a given angle .
In order to use such a spectrometer in the von Hamos configuration even with a diffused source, the idea is to exploit a situation like the one pictured in fig. \ref{compare_big}e.
In figure, signal photons coming with the right Bragg angle are shown in yellow while the background photons are shown in red.
If the slits widths are set in order to have, for a specific beam divergence, the origin of the beam cone in an imaginary point behind the real extended source, 
then X rays coming from a bigger portion of the target will imping on the HAPG with the proper angle, resulting in a higher statistics. 
With a proper selection of the slits positions and widths (black), also the background due to the X rays produced in the wrong position but with the correct Bragg angle can be kept under control.
This will be the main subject of the further stages of the VOXES project.

\section{The VOXES project}

\begin{figure}[htb]
\centerline{%
\includegraphics[width=11cm]{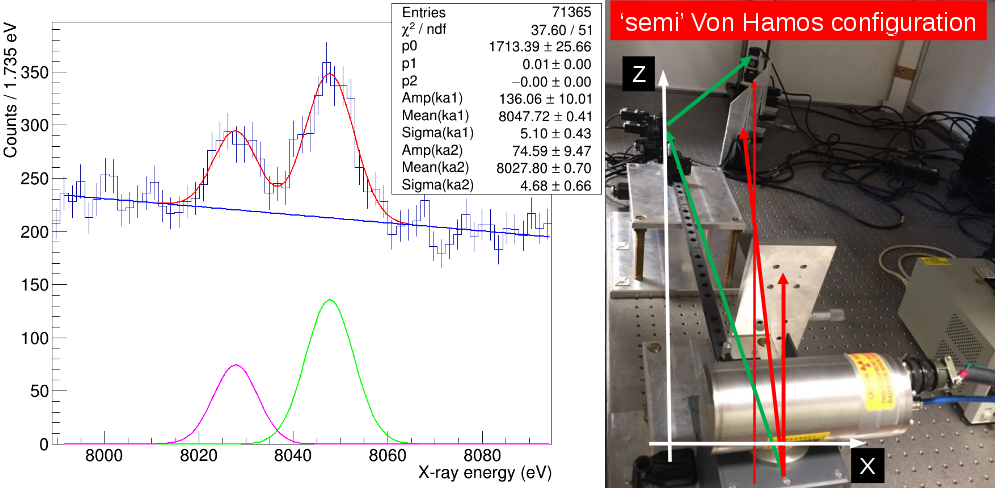}}
\caption{\em Left: fitted Bragg spectrum of Cu $K_{\alpha 1}$ and $K_{alpha 2}$ lines. Right: picture of the setup used for the measurement.}
\label{spectra}
\end{figure}

The VOXES project aims to deliver a cost effective system having an energy resolution at the level of eV for X-ray energies from about 2 keV up to tens of keV, 
able to work also with diffused sources. To reach this goal, a Bragg spectrometer in both the von Hamos and Johann geometry, 
using HAPG crystals combined with precision position detectors is presently under investigation. 
The first measurement with the VOXES spectrometer has been carried out at the INFN laboratories of Frascati (LNF) using the ``semi''-von Hamos geometry, 
where an X-ray tube was used to activate a $2 \,mm$ thick Cu foil, and it is presented in this section.
The reflecting crystal is cylindrically shaped with a radius
of curvature of $206.7\,mm$ and dimensions of $30\,mm\times32\,mm$; the substrate is a concave N-BK7 glass lens (Thorlab LK1487L1), 
coated with a single $100\,\mu m$ HAPG layer by the company Optigraph (Berlin, Germany).
Given a declared mosaicity of $0.007^{\circ}-0.01^{\circ}$, the isotropically emitted $K_{\alpha 1}$ and $K_{\alpha 2}$ X-ray lines have been sorted by two adjustable slits (Thorlab VA100/M) in order to 
keep a beam divergence of $\Delta\theta = 0.1^{\circ}$ and to have a point-like source spot.  A picture of the setup is shown in fig. \ref{spectra}, right. 
In order to match the ``semi'' von Hamos configuration,
both HAPG source-to-crystal and crystal-to-detector distances are $900,54\,mm$ in the XZ dispersion plane; 
Bragg angle ($\theta_{B}=13,29^{\circ}$) is tuned to match the average energy between the two lines.
The  two slits, S1 ($175\,\mu m$ width) and S2 ($1.11\, mm$ width) are placed $100 \,mm$ and $635 \,mm$ after the source, respectively; 
source position is defined as the center of the Cu foil.  
the resulting horizontal beam size on the HAPG crystal is $1,57\,mm$ when directly facing the beam direction and $1,62\,mm$ when rotated of $\theta_{B}$.
The position detector is a commercial MYTHEN2-1D 640 channels strip detector produced by Dectris (Zurich, Switzerland); the active area is $32\times8\,mm^2$,
 strip width and thickness are, respectively, $50\,\mu m$ and $420\,\mu m$. The calibration is done using the two known position of Cu $K_{\alpha 1}$ and $K_{\alpha 2}$.
To avoid direct shining of the primary X-rays on the MYTHEN2-1D surface, an aluminum structure and three $2\,mm$ thick PVC foils have been put on the unwanted path (see fig. \ref{compare_big}e).
The measured spectrum is a result of a 12 hours of data taking and it is shown in fig. \ref{spectra}, left.

\noindent The fit, performed with two gaussians and a second degree polynomial, shows a very promising resolution of $5.1\, eV$ ($\sigma(K_{\alpha 1})$) and $4.7\, eV$ ($\sigma(K_{\alpha 2})$), 
allowing a $0.4\,eV$ and $0.7\,eV$ precision measurement of the two lines respectively.

\section{Conclusions}

In the framework of the VOXES project, 
a first measurement of the Cu characteristic $K_{\alpha 1}$ and $K_{\alpha 2}$ X-ray lines has been presented, 
showing a very promising resolution of $5.1\, eV$ ($K_{\alpha 1}$) and $4.7\, eV$ ($K_{\alpha 2}$), 
allowing a $0.4\,eV$ and $0.7\,eV$ precision measurement of the two lines respectively. 
This results will be improved, thanks to a system of motorized rotators and positioners, by fine tuning of the Bragg angle, the XYZ position of both the HAPG crystal and MYTHEN2-1D detector and 
tilt adjustment of the HAPG; in the future, a series of measurements to estimate the efficiency of such a spectrometer and to explore the possibility to use it also
with non point-like sources varying the slits apertures and positions are also planned. The performances of the same HAPG+MYTHEN2D-1D system in the Johann configuration will be also investigated.\\

\section*{Acknowledgements}
\emph{This work is supported by the 5th National Scientific Committee of INFN in the framework of the Young Researcher Grant 2015, n. 17367/2015.
We thank to the LNF staff and in particular to the SPCM service for the support in the preparation of the setup.}

\end{document}